\documentclass{cimento}

\usepackage{graphicx}  
\usepackage{siunitx}
\usepackage{mhchem}

\title{Triple-GEM detectors for the Phase-2 upgrade of the CMS muon spectrometer}
\author{A.~Pellecchia\from{ins:x}\ETC ~on behalf of the CMS Muon group}
\instlist{\inst{ins:x} INFN, Sezione di Bari - Bari, Italy}

\begin{document}

\maketitle

\begin{abstract}
The High-Luminosity LHC will deliver an unprecedented instantaneous luminosity, requiring all experiments to upgrade their detectors to sustain the higher background rates.
The upgrade of the CMS Muon spectrometer includes three stations of triple-GEM detectors.
We present the status of the commissioning and performance validation of the three GEM stations:
GE1/1, which is taking data in Run 3, GE2/1, for which the first detectors will be installed in 2024, and ME0, which has undergone several performance studies for high-rate and longevity and will start its mass production in 2024.
\end{abstract}

\section{Introduction}

\begin{figure}[tb]
	   \centering
	   \includegraphics[width=.8\textwidth]{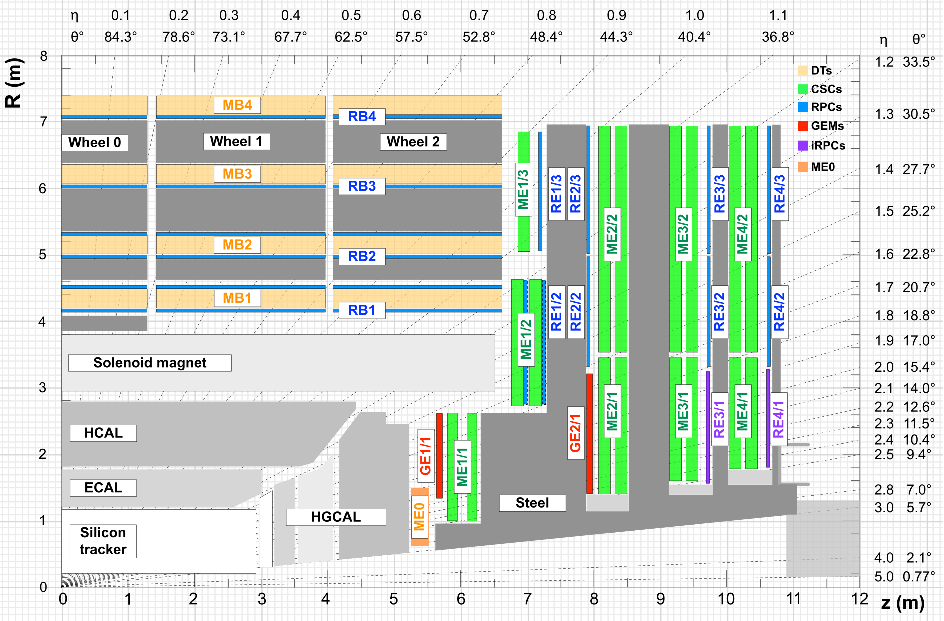}
	   \caption{Quadrant of a section of the CMS muon spectrometer, showing the Phase-1 muon stations and the new stations after the Phase-2 upgrade (in red and orange) \cite{ref:muon_tdr}.}
	   \label{fig:cms_muons}
   \end{figure}

   The High-Luminosity LHC will deliver a peak luminosity of \SI{5e34}{\per\centi\m\per\s}, requiring the Compact Muon Solenoid (CMS) experiment \cite{ref:cms} to upgrade its detectors to maintain a good trigger performance in the increased pile-up.
   The CMS muon spectrometer will be upgraded with new front-end and trigger electronics and the addition of new detector stations of resistive plate chambers (RPC) and gas electron multipliers (GEM) \cite{ref:muon_tdr}.
   The CMS GEM upgrade will consist of three detector stations in the endcap region, GE1/1, GE2/1 and ME0 (Fig.\,\ref{fig:cms_muons}), to complement the existing cathode-strip chambers (CSC) in the high pseudorapidity region and extend the geometrical acceptance of the muon system in the very forward region.

   Each of the three stations is made of a disk of trapezoidal triple-GEM detectors in both endcaps.
   The GE1/1 station has been installed in the second LHC long shutdown (LS2), commissioned and operated in the first years of Run 3;
   ME0 will be installed in CMS in the beginning of LS3, while the GE2/1 installation is planned to be divided among the LS3 and shorter LHC technical stops.
   The three stations share a common detector design, gas mixture (Ar/\ce{CO2} 70\%/30\%) and front-end electronics.

\section{GE1/1 system performance}

The GE1/1 station will complement the ME1/1 CSC station, extending the lever arm for the $p_T$ measurement in the pseudorapidity region $1.6 < |\eta| < 2.15$ and allowing lower Level-1 trigger rates \cite{ref:gem_tdr}.

The GE1/1 system is made of 36 super-chambers (SC) per endcap, each made of two stacked trapezoidal triple-GEM detectors with an aperture of 10°;
to decrease the discharge energy, the GEM foils are segmented along the $\eta$ direction on the top side in 47 sectors for even-numbered superchambers and 40 sectors for odd-numbered ones and each sector is connected to a common high voltage pad through a decoupling protection resistor of \SI{10}{\mega\ohm}.

The installation of GE1/1 occurred between 2019 and 2020;
after installation, the GE1/1 detectors were operated at first during LS2 in the CMS cosmic runs at zero tesla (CRUZET) and with the CMS magnet turned on at its full field of four tesla (CRAFT), as well as during Run 3 in 2022 and 2023.

During the first ramping phases of the CMS magnet, a high number of discharges in the GEM system were observed;
these observations, together with additional studies performed with spare GE1/1 detectors in the Goliath magnet at the CERN North Area, prompted the definition of a special operation mode of the detectors during magnet ramps, consisting of powering the GEM foils at a voltage difference of \SI{430}{\volt} while keeping the gaps off, to reduce the probability of discharge damage to the GEM foils \cite{ref:goliath}.

\begin{figure}[tb]
	   \centering
	   \includegraphics[width=\textwidth]{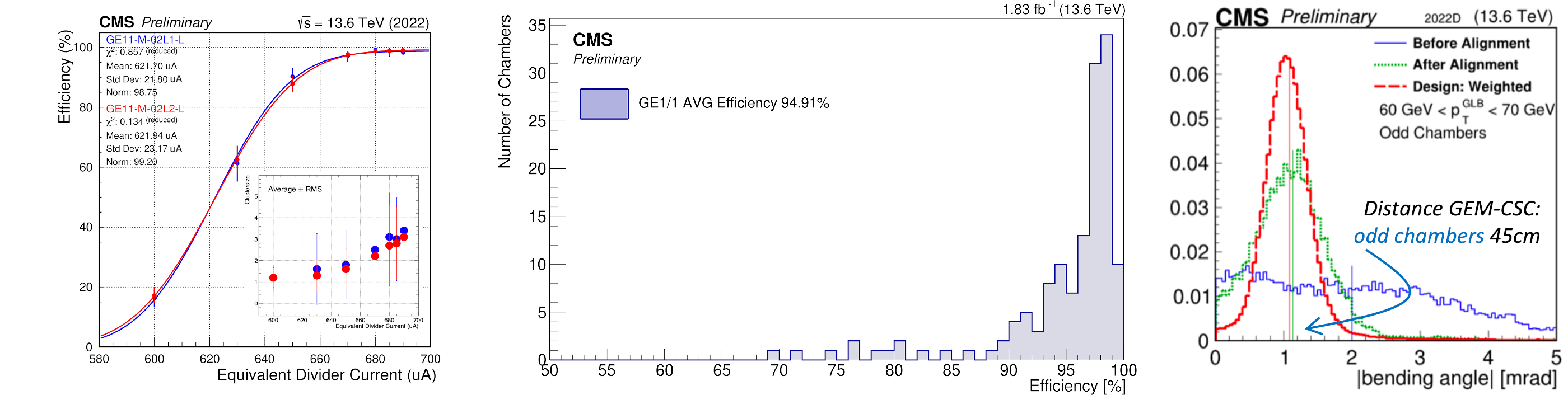}
	   \caption{
			(a) Efficiency of a GE1/1 detector in CMS measured with muons from pp collisions as a function of its HV point.
			(b) Distribution of the efficiencies of all GE1/1 detectors in CMS \cite{ref:frengo}.
			(c) Bending angle measurement between GE1/1 and ME1/1 segment \cite{ref:devin}.
			}
	   \label{fig:ge11_efficiency}
   \end{figure}

The detector efficiency was measured using cosmics and muon segments from the ME1/1 station during the initial proton-proton collisions of Run 3 (Fig.\,\ref{fig:ge11_efficiency}a).
During 2023 an efficiency measurement as a function of the HV point was performed on all detectors;
the result (Fig.\,\ref{fig:ge11_efficiency}b) show a plateau efficiency higher than the design requirements of 97\%, while a minority of the detectors show localized areas of inefficiency, due mainly to two effects \cite{ref:frengo}.

The first inefficiency source is the loss of communication of front-end electronics due to the decreasing receiver signal strength of the VTRx transceiver \cite{ref:vtrx};
the second is the presence of short circuits -- a possible consequence of discharge -- in one of the GEM foil sectors.
At the end of the LHC activity in 2023 there were 37 short circuits in the GE1/1 system, each rendering inactive 1/40 of the detector area.
The treatment of the short circuit with an on-site cleaning procedure is a possible mitigation strategy, for which the effectiveness is being evaluated.

Following the efficiency measurement, the HV point of each electrode has been set to the plateau point for efficiency and lowered where needed to reduced the occurrence of discharges after a dedicated analysis \cite{ref:simone}.

A software alignment with respect to the neighbouring ME1/1 CSC station has been performed from an analysis of the tracking data;
after alignment, a measurement of the muon bending angle between the GE1/1 and ME1/1 segments has been possible (Fig.\,\ref{fig:ge11_efficiency}c).
Implementation of the alignment and the benging angle measurement in the Level-1 trigger, needed to complete the GE1/1 trigger integration, are ongoing.

\section{GE2/1 detector performance}

The GE2/1 system will complement the ME2/1 station, covering the pseudorapidity region $1.6 < |\eta| < 2.4$;
it will also allow triggering at Level-1 on displaced vertices, which will benefit searches in physics channels including neutral long-lived particles.

The design of each GE2/1 detector is similar to GE1/1, with the difference that each chamber is made of four smaller detector modules along the $\eta$ direction.
Moreover, design improvements have been introduced following the operational experience of GE2/1:
\begin{itemize}
    \item in GE2/1 detectors, the first and second GEM foils are segmented also on the bottom side to improve the discharge stability;
    \item in GE1/1, possible mechanical deformations in the drift and readout PCBs reduced the gain and timing uniformity; for this reason, in GE2/1 detectors the PCBs are held by internal pillars in the center of the active area;
    \item the interface with the detector read-out strips and the readout electronics was redesigned for a more redundant grounding to reach lower noise.
\end{itemize}

The GE2/1 detector performance has been measured on a pre-production detector in two test beam campaigns carried out at the CERN North Area with high-energy muons in 2021 and 2022 \cite{ref:gem_tb};
the space resolution was measured with tracks built by a high-resolution triple-GEM tracker and found equal to \SI{346+-6}{\micro\radian}, in agreement with the design requirement of less than \SI{500}{\micro\radian}.
The efficiency of the GE2/1 detector over an area of $10\times\SI{10}{\centi\m\squared}$ was measured with high granularity (Fig.\,\ref{fig:ge21_me0}a) and found to be locally higher than 99\% on the majority of the surface;
narrow inefficiency regions with a width between 180 and \SI{220}{\micro\m}, with dips higher than 50\%, were found in the separation between two GEM foil sectors, limiting the average efficiency over the irradiated area to 98\%.


A GE2/1 demonstrator was installed in CMS in the 2021 LHC year-end technical stop (YETS) and integrated with the CMS data acquisition and control systems.
The installation of the full system is expected to begin in early 2024 with two detectors and to continue during the future LHC technical stops and LS3.

\section{ME0 detector optimization and performance}

The ME0 system will cover the very forward region of the CMS Muon system, complementing the other GEM and CSC stations in the pseudorapidity region $2 < |\eta| < 2.4$ and extending the Muon system acceptance in the region $2.4 < |\eta| < 2.8$;
this extension will allow increasing the CMS sensitivity for forward decay processes such as $H\to ZZ\to \mu\mu ll$.
In order to operate standalone in the reconstruction of muon segments at the Level-1 trigger, ME0 will be made of stacks of six triple-GEM detectors, each covering an angle of 20°.

The ME0 detector design benefits of the improvements already implemented for GE1/1 and GE2/1;
moreover, the expected background particle rate in the highest pseudorapidity region of ME0 (\SI{150}{\kilo\Hz/\centi\m\squared}) is unprecedented for a large-area gaseous detector;
consequently, dedicated studies have been performed for ensuring good operations in high rates and high integrated charges.

The ageing of a triple-GEM detector has been measured by monitoring its effective gain under x-ray irradiation up to a integrated charge of \SI{8}{\coulomb/\centi\m\squared}, equal to the expected integrated charge after 10 years of ME0 operations in HL-LHC;
no performance degradation was observed \cite{ref:ageing}.

Measurements carried out in laboratory with x-rays \cite{ref:rate_capability} and in the gamma irradiation facility (GIF++) at CERN \cite{ref:gif} have set the rate capability of a triple-GEM detector with the ME0 design to \SI{10}{\mega\Hz} per GEM foil sector, independently of the sector area;
this limitation is due to the voltage drops on the protection resistors due to the avalanche currents collected by the GEM foils under irradiation.
The nominal gain can be recovered by applying an overvoltage ``compensation''.
From the results of the rate capability studies, the value of the protection resistor for the ME0 GEM foils has been set to \SI{2}{\mega\ohm} and the foil segmentation pattern has been adapted to the radial direction, to allow an equal gain drop across all the sectors and making the gain compensation feasible using a single high-voltage channel per electrode.

Two ME0 detector prototypes in their final design have been operated in test beam in 2021 and 2022, showing a space resolution of \SI{235+-2}{\micro\radian}, in agreement with the design requirement, and an efficiency locally higher than 99\%.
Due to the higher number of sectors in a smaller area compared with GE2/1, the foil segmentation introduces a higher number of dead areas and reduces the average detector efficiency to 95\%.
An innovative GEM foil manufacturing technique, called ``random hole segmentation'', has been tested in an ME0 prototype and found to lower the efficiency loss in presence of segmentation \cite{ref:segmentation}, but has not been adopted for mass production as it is still missing validation studies over sustained operation, which would not be compatible with the ME0 delivery schedule.

\begin{figure}[tb]
	\centering
	\includegraphics[width=\textwidth]{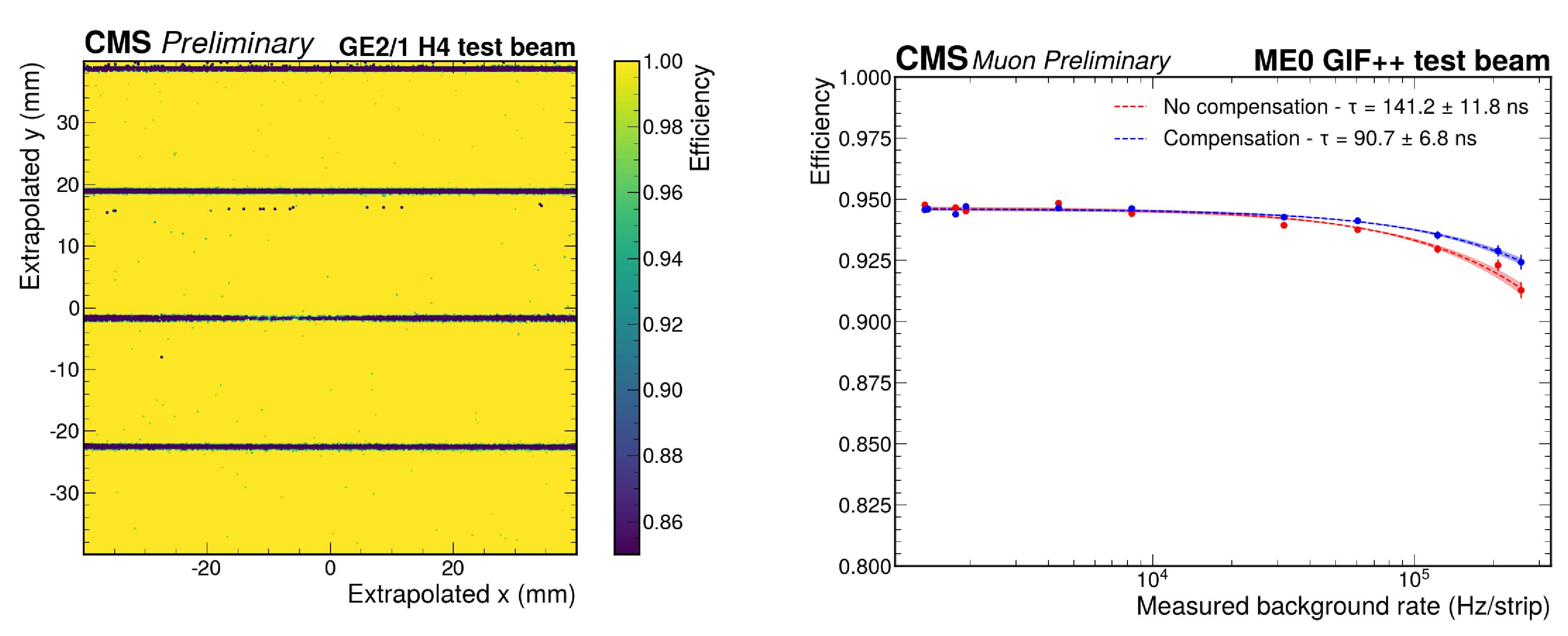}
    \caption{(a) Efficiency map of a GE2/1 detector measured at test beam with 80 GeV/c muons. (b) Efficiency of an ME0 detector for 80 GeV/c muons measured in test beam at the GIF++ as a function of the background rate.}
	\label{fig:ge21_me0}
\end{figure}

In 2022, an ME0 detector has been operated in a test beam with high-energy muons and high-background irradiation at the GIF++, with the goal of measuring the detector efficiency in presence of background with final electronics.
The test beam results have shown an efficiency loss of about 2.5\% under irradiation of \SI{200}{kHz}/strip due to the dead time of the front-end electronics, measured to be \SI{200}{\nano\s} (Fig.\,\ref{fig:ge21_me0}b);
the efficiency drop is expected to be mitigated by the redundancy provided by the six layers per stack, with an efficiency drop on the entire stack of the order of 1\%, and does not require changes to the detector design.

\section{Conclusions}

The GEM project consists of three stations of triple-GEM detectors to complement and extend the existing muon spectrometer of the CMS system.
The GE1/1 station, produced and installed in CMS, has been commissioned during the CMS cosmic runs and Run 3 and its configuration has been optimized for performance and stability.
The performance of the detectors for the GE2/1 and ME0 stations have been verified in test beam;
for ME0, dedicated studies have been focused on longevity and rate capability, resulting in design optimizations to sustain high rates and implement a dynamic compensation of gain drop.
The main ongoing efforts are directed towards the ME0 stack integration and timing optimization, in view of the mass production starting in 2024.


\end{document}